\documentstyle[epsfig]{mn2e}

\title[The coevolution of Pops II and III]{The First Billion Years project - III:  The impact of stellar radiation on the coevolution of Populations II and III}

\author[J. L. Johnson, C. Dalla Vecchia and S. Khochfar]
       {Jarrett L. Johnson$^1$$^,$$^2$\thanks{E-mail: jlj@lanl.gov}, Claudio Dalla Vecchia$^2$ 
          and Sadegh Khochfar$^2$ \\
 $^1$Los Alamos National Laboratory, Los Alamos, NM 87545, USA \\
Nuclear and Particle Physics, Astrophysics and Cosmology Group (T-2) \\  
 $^2$Max-Planck-Institut f{\"u}r extraterrestrische Physik, Giessenbachstra\ss{}e, 85748 Garching, Germany \\
Theoretical Modeling of Cosmic Structures Group}

\pubyear{2012}

\begin{document}
\maketitle
\topmargin-1.2cm
\addtolength{\oddsidemargin}{+.5cm}
\addtolength{\evensidemargin}{+.5cm}

\begin{abstract}
With the first metal enrichment by Population (Pop)~III supernovae (SNe), the formation of
the first metal-enriched, Pop~II stars becomes possible.  In turn, Pop~III star formation and  
early metal enrichment are slowed by the high energy radiation emitted by Pop~II stars.  
Thus, through the SNe and radiation they produce, Populations
II and III coevolve in the early Universe, one regulated by the other.  We present large (4 Mpc)$^3$, 
high resolution cosmological simulations in which we self-consistently model early metal enrichment and the stellar
radiation responsible for the destruction of the coolants (H$_{\rm 2}$ and HD) required
for Pop~III star formation.  We find that the molecule-dissociating stellar radiation produced both locally 
and over cosmological distances reduces the Pop~III star formation rate at $z$ $\ga$ 10 by up to an 
order of magnitude, to a rate per comoving volume of $\la$ 10$^{-4}$ M$_{\odot}$ yr$^{-1}$ Mpc$^{-3}$, 
compared to the case in which this radiation is not included.  
However, we find that the effect of LW feedback is to enhance the amount of Pop~II star formation. 
We attribute this to the reduced rate at which gas is blown out of dark matter haloes by SNe in the simulation 
with LW feedback, which results in larger reservoirs for metal-enriched star formation.
Even accounting for metal enrichment, molecule-dissociating radiation and the strong suppression of low-mass galaxy formation 
due to reionization at $z$ $\la$ 10, we find that Pop~III stars
are still formed at a rate of $\sim$ 10$^{-5}$ M$_{\odot}$ yr$^{-1}$ Mpc$^{-3}$ down to $z$ $\sim$ 6.  This 
suggests that the majority of primordial pair-instability SNe that may be uncovered in future surveys
will be found at $z$ $\la$ 10.
We also find that the molecule-dissociating radiation emitted from Pop~II stars 
may destroy H$_{\rm 2}$ molecules at a high enough rate to suppress gas cooling and allow 
for the formation of supermassive primordial stars which collapse to form $\sim$ 10$^5$ M$_{\odot}$ black holes.
\end{abstract}


\section{Introduction}
The gravitational collapse of primordial gas into the first, Population (Pop)~III stars at $z$ $\ga$ 20 marks the end of the 
cosmic dark ages and ushers in an era of rapidly increasing complexity in the early Universe 
(e.g. Barkana \& Loeb 2001; Ciardi \& Ferrara 2005).  Expected to be typically
much more massive than most stars formed today (e.g. Bromm \& Larson 2004; Glover 2005), the first stars emit
copious high energy radiation that ionizes and heats the gas in and around their host dark matter (DM) haloes
(e.g. Kitayama et al. 2004; Whalen et al. 2004; Alvarez et al. 2006; Abel et al. 2007; Johnson et al. 2007), and
also destroys H$_{\rm 2}$ and HD molecules in the primordial gas over cosmological distancees (e.g. Haiman et al. 1997; Glover \& Brand 2001;
Ahn et al. 2009; Holzbauer \& Furlanetto 2012).  As these molecules are critical for the cooling of the primordial gas, 
it is expected that their destruction leads to diminished rates of gas collapse and Pop~III star formation (e.g. 
Omukai \& Nishi 1999; Machacek et al. 2001; Mesinger et al. 2006a; Wise \& Abel 2007; O'Shea \& Norman 2008; 
Trenti et al. 2009).

At the end of their brief (10$^6$ - 10$^7$ yr) lives, a large fraction (e.g. Heger et al. 2003) of Pop~III stars 
explode as supernovae (SNe) and eject the first heavy elements into the intergalactic medium (e.g. Ferrara et al. 2000; 
Bromm et al. 2003; Kitayama \& Yoshida 2005; Greif et al. 2007; Vasiliev et al. 2008; Whalen et al. 2008).
This sets the stage for the onset of metal-enriched, Pop~II star formation (e.g. Wise \& Abel 2008; Greif et al. 2010; Maio et al. 2011)
in the first galaxies (e.g. Bromm \& Yoshida 2011).  The lower surface temperatures of Pop~II stars compared to Pop~III stars
lead to a lower efficiency (per stellar mass) 
of ionizing photon production (e.g. Oh et al. 2001; Tumlinson et al. 2001; Schaerer 2002), but also lead to 
only a modest decrease in the efficiency of molecule-dissociating, Lyman-Werner (LW) photon production (e.g. Greif \& Bromm 2006), 
as well as to an increased efficiency in the production of H$^-$-ionizing photons (e.g. Shang et al. 2010).\footnote{Because molecular hydrogen is formed in the primordial gas largely via the reaction H$^-$ + H $\to$ H$_{\rm 2}$ + e$^-$, the destruction of H$^-$ effectively slows the production of H$_{\rm 2}$ (see e.g. Chuzhoy et al. 2007).}  Thus, the high energy radiation emitted from early generations of 
Pop~II stars can have a dramatic impact in slowing the rate of Pop~III star formation.

This interplay between Pops~II and III star formation constitutes a feedback loop whereby
Pop~II star formation can only take place in regions enriched by Pop~III stars and the pace of Pop~III star formation 
(and the subsequent metal enrichment) is regulated by the amount of radiation emitted by Pop~II stars.  
Therefore, in order to properly model the earliest episodes of star and galaxy formation, it is necessary to model the formation 
of both populations and their respective chemical and radiative feedback, as a coupled system.

Much previous work has treated these processes, with many results gleaned from cosmological simulations of early SNe feedback and metal enrichment (e.g. Tornatore et al. 2007; Wise \& Abel 2008; Wiersma et al. 2009a; Greif et al. 2010; Maio et al. 2011; Wise et al. 2012) and of the build-up of the global background (e.g. Yoshida et al. 2003; Wise \& Abel 2005; Johnson et al 2008) or the locally generated (e.g. Dijkstra et al. 2008; Ahn et al. 2009; Hummel et al. 2011; Petkova \& Maio 2011; Wise et al. 2012) 
stellar LW radiation field.  
Previous authors have also modelled the impact of the LW background on star formation, both self-consistently (Ricotti et al. 2002; 2008; Trenti et al. 2009; Petri et al. 2012) and at fixed levels (e.g. Kuhlen et al. 2012; Safranek-Shrader et al. 2012), together with simplified treatments of early metal enrichment.  Assembling all of these ingredients -- star formation, mechanical feedback and metal enrichment from SNe, and both local and cosmological stellar LW radiation fields -- self-consistently in a cosmological context is the next step in simulating the formation of the earliest galaxies.

Here we present the results of cosmological simulations which accomplish this task, self-consistently accounting for
early metal enrichment and mechanical feedback from Pop~II and Pop~III SNe, as well as the impact of 
both the locally-generated and the cosmological background stellar LW radiation from both populations.  
While our simulation is of high enough resolution to track even the first episodes of star formation in minihaloes, 
we simulate a relatively large cosmological volume in order to follow the 
assembly of galaxies down to $z$ $\simeq$ 6.  Thus, as we model galaxy formation in detail 
from the epoch of the first stars through the entire epoch of reionization, our results offer arguably 
the most complete picture to date of galaxy formation in the early Universe.
    
In the next Section, we begin by describing the simulations that we have carried out, with particular
attention paid to our implementation of LW feedback.  In Section 3 we present our results, highlighting the 
impact that LW radiation has on star formation and chemical enrichment.  Finally, we give our conclusions
and provide a brief discussion of our results in Section 4.

\section{The simulations}
The simulations we have carried out are two in a larger series of simulations that constitute 
the {\it First Billion Years} (FiBY) project (Khochfar et al. 2012 in prep).
While we leave many of the details of the simulations to be described elsewhere (Dalla Vecchia et al. in preparation), 
here we highlight the main aspects of the star formation (Section 2.1), metal enrichment (Section 2.3), and reionization (Section 2.4)
prescriptions that we have employed.  As the simulations presented here are unique in the full suite of 
FiBY simulations in that they include LW feedback from stars, we present our implementation of this effect 
in detail in Section 2.2.   

For all of our simulations we use a modified version of the smoothed-particle hydrodynamics (SPH) code GADGET (Springel et al. 2001; Springel 2005) that was previously developed in the {\it Overwhelmingly Large Simulations} (OWLS) project (Schaye et al. 2010).  Among the modifications is line cooling in photoionization equilibrium for 11 elements (H, He, C, N, O, Ne, Mg, Si, S, Ca, Fe) following Wiersma et al. (2009b), which is computed with CLOUDY v07.02 (Ferland 2000), as well as prescriptions for SNe mechanical feedback and metal enrichment
as described in Section 2.3.  For the FiBY project, we have furthermore implemented a full non-equilibrium primordial chemistry network and
molecular cooling functions for both H$_{\rm 2}$ and HD, following Abel et al. (1997), Galli \& Palla (1998), Yoshida et al. (2006), and Maio et al. (2007).  Along with this, we have developed prescriptions for Pop~III stellar evolution and chemical feedback which track the enrichment of the gas
in each of the 11 elements listed above individually, following the stellar yields provided by Heger \& Woosley (2002, 2010).

We carry out two simulations, one with and one without LW feedback from stars, using identical cosmological (periodic) initial conditions
within a cubic volume 4 Mpc (comoving) on a side.  We include both DM and gas, with an SPH particle mass of 1.25 $\times$ 10$^3$ M$_{\odot}$ 
and a DM particle mass of 6.16 $\times$ 10$^3$ M$_{\odot}$.  The simulation is initiated with 684$^3$ SPH particles and an
identical number of DM particles.
Our assumed cosmological parameters are consistent with the results reported by the {\it Wilkinson Microwave Anisotropy Probe} (WMAP) team (e.g. Komatsu et al. 2009):
$\Omega_{\rm m}$ = 0.265, $\Omega_{\rm b}$ = 0.0448, $\Omega_{\rm \Lambda}$ = 0.735, $H_{\rm 0}$ = 71 km s$^{-1}$ Mpc$^{-1}$, and $\sigma_{\rm 8}$ = 0.81.

\subsection{Star formation in the FiBY}
Our prescription for star formation is based on a pressure law and is designed to yield results 
consistent with the observed Schmidt-Kennicutt law (Schmidt 1959; Kennicutt 1998), as described by
Schaye \& Dalla Vecchia (2008).  
We set the threshold density above which star formation occurs to $n$ = 10 cm$^{-3}$ which, as we argue in 
Appendix A, is sufficiently high to resolve the impact that LW radiation has in destroying 
H$_{\rm 2}$ molecules in the primordial gas and diminishing the efficiency with which it cools.  As the
Pop~III star formation rate (SFR) is strongly regulated by this feedback, it is crucial that we resolve such high densities.
Above this threshold density, we use an effective equation of state with the pressure (over Boltzmann's constant $k_{\rm B}$) normalized to 
$P_{\rm 0}$/$k_{\rm B}$ = 10$^2$ cm$^{-3}$ K at $n$ =10 cm$^{-3}$, with an effective adiabatic 
index $\gamma_{\rm eff}$ =4/3.  More details about the star formation prescription, and the motivation for it, 
can be found in Schaye \& Dalla Vecchia (2008).

Below we describe our choices for the IMFs of Pop~II and III stars, as well as our choice for the 
metallicity at which the IMF changes from one to the other.

\subsubsection{Pop~III star formation}
As the properties of metal-free stellar populations are expected to be markedly different from those
of metal-enriched populations, both due to the lower opacity of their interiors (e.g. Siess et al. 2002; Lawlor et al. 2008) which leads to 
higher surface temperatures (e.g. Schaerer 2002) and to a more 
top-heavy initial mass function (IMF; e.g. Bromm \& Larson 2004), our star formation prescriptions 
differ depending on the metallicity of the star-forming gas.  

For the IMF of Pop~III stars, we assume that it takes the form of a power-law with a Salpeter slope (Salpeter 1955).  While the 
Pop~III IMF is unknown and not well-constrained, this functional form is found to describe well the present-day IMF in a 
large variety of environments (e.g. Bastian et al. 2010; see however Cappellari et al. 2012 on a systematic changing IMF).  
Along with this, the upper ($M_{\rm *, upper}$) and lower ($M_{\rm *, lower}$)  initial mass limits of stars fully specify the IMF.  
For our choice of these values, we follow Bromm \& Loeb (2004) and Karlsson et al. (2008) for 
the upper and lower mass limits, respectively.  We choose $M_{\rm *, upper}$ = 500 M$_{\odot}$, as the former authors 
show that this is roughly the maximum mass that is likely attainable for Pop~III stars under typical conditions (see also Omukai \& Palla 2001, 2003).  The latter authors
used data on the chemical abundances of metal-poor stars to estimate that only a fraction $\gamma_{\rm PISN}$ $\la$ 0.07 of Pop~III stars had masses 
in the range (between $\simeq$ 140 and $\simeq$ 260 M$_{\odot}$; e.g. Heger et al. 2003) in which they explode as powerful pair-instability SNe (PISNe). 
For the lower mass limit we assume $M_{\rm *, lower}$ $=$ 21 M$_{\odot}$, which 
is the maximum value consistent with this limit on $\gamma_{\rm PISN}$ and
with our assumption of a Salpeter slope. Finally, we note that recent cosmological simulations which have modelled
the radiative feedback from Pop~III stars suggest that they may typically form with masses within the range we have chosen (see e.g. Hosokawa et al. 2011; Stacy et al. 2012).

\subsubsection{Pop~II star formation}
For the Pop~II stellar IMF, we adopt that of Chabrier (2003), which has a Salpeter slope at the high mass
end and which extends down to sub-solar masses.  This choice is motivated by theoretical studies which 
show that, with increasing metallicity of the star-forming gas, the IMF shifts rapidly from that of Pop~III
stars to a much more bottom-heavy IMF, with characteristic stellar masses much like those in the present-day
Universe (e.g. Bromm et al. 2001; Schneider et al. 2003; Omukai et al. 2005; Santoro \& Shull 2006; Dopcke et al. 2011, 2012).

While the 'critical metallicity' at which this transition in the stellar IMF takes place is unknown,
we have chosen a value of the metallicity above which Pop~II stars form and below which
Pop~III stars form\footnote{We note that while, by definition, any star formed with a non-zero metallicity is not a Pop~III star,
 we refer to Pop~III stars here as those formed with a metallicity lower than the critical metallicity
required for the IMF to transition from that common to Pop~III stars to that inferred for 
stars in the present-day Universe.  We emphasize that it is still possible that some low-mass ($\la$ 1 M$_{\odot}$)
stars may form from primordial gas, although the overall IMF of primordial stars is expected to 
be very top-heavy relative to that of Pop~II stars (e.g. Greif et al. 2011).} that is roughly consistent with the prevailing theory as well as 
with the inferred metallicities of the most metal-poor stars (e.g. Frebel et al. 2007; Caffau et al. 2011).  
The critical metallicity that we assume is $Z_{\rm crit}$ = 10$^{-4}$ Z$_{\odot}$ (with $Z_{\odot}$ = 0.02).
Based on the results of similar cosmological simulations (e.g. Maio et al. 2011), we do not expect our 
results to depend sensitively on this choice.  In a companion paper, the details of the Pop~III to Pop~II
transition in the FiBY simulations will be discussed further (Maio et al. in prep).

\subsection{H$_{\rm 2}$ photodissociation and H$^{-}$ photodetachment feedback from stars in the FiBY}
In our simulation including LW feedback from stars, we include the effects of two photoreactions which impact the abundance of molecular hydrogen (H$_{\rm 2}$),  H$_{\rm 2}$ dissociation and H$^{-}$ detachment:

\begin{equation}
{\rm H}_{\rm 2} + h\nu \to 2{\rm H}  \mbox{\ } \nonumber 
\end{equation}
\begin{equation}
{\rm H}^{-} + h\nu \to {\rm H} + {\rm e}^{-}  \mbox{\ } \mbox{\ .}
\end{equation}

We consider separately the cosmological background LW radiation field from distant sources and that produced by local sources.  These
components of the dissociating and ionizing flux will be added together to estimate the time- and space-dependent total radiative flux.  In turn, these fluxes are converted into photodestruction rates, which are updated on the fly in our primordial chemical network as described below in Section 2.2.4.

\subsubsection{The contribution from a radiation background}
As the mean free path of LW photons in the early Universe can be up to $\sim$ 10 physical Mpc (e.g. Haiman et al. 1997) -- much larger scales than are captured in our 
simulation volume -- we must account for the presence of a cosmological LW background radiation field generated by stars well outside
our simulation box.  To obtain an estimate of the level of the LW background, we follow the approach outlined by Greif \& Bromm (2006) which accounts for the cosmological rates of both Pop II and Pop III star formation.  In general, we have for the flux $J_{\rm LW}$ of LW radiation, as a function of the the mass density 
in stars $\rho_{\rm *}$ and the number of LW photons produced per stellar baryon $\eta_{\rm LW}$, 

\begin{equation}
J_{\rm LW} \simeq \frac{h c}{4 \pi m_{\rm H}} \eta_{\rm LW} \rho_{\rm *} \left(1 + z \right)^3 \mbox{\ ,}
\end{equation}
where $c$ is the speed of light, $h$ is Planck's constant, $m_{\rm H}$ is the mass of the hydrogen atom, and $z$ is the redshift.
With this we then find for the flux of the LW background (in units of 10$^{-21}$ erg s$^{-1}$ cm$^{-2}$ Hz$^{-1}$ sr$^{-1}$), as a 
function of the Pop~III and Pop~II  SFRs, repsectively,

\begin{equation}
J_{\rm 21, III, bg} \simeq 1.5 \left(\frac{1+z}{16}\right)^{3} \left(\frac{SFR_{\rm III}}{10^{-3} {\rm M_{\odot} \, yr^{-1} \, Mpc^{-3}}}\right)  \mbox{\ ,}
\end{equation}
\begin{equation}
J_{\rm 21, II, bg} \simeq 0.3 \left(\frac{1+z}{16}\right)^{3} \left(\frac{SFR_{\rm II}}{10^{-3} {\rm M_{\odot} \, yr^{-1} \, Mpc^{-3}}}\right)  \mbox{\ ,}
\end{equation}
where the SFRs are defined as those per comoving volume.
To arrive at these formulae we have assumed that the stars producing the bulk of the LW photons, for 
both Pop III and Pop II, live only 5 Myr (e.g. Schaerer 2002; Leitherer 1999); also we have assumed $\eta_{\rm LW}$ 
to be given by the values adopted by Greif \& Bromm (2006) for Pop III stars.\footnote{We have taken the value of $\eta_{\rm LW}$ = 2 $\times$ 10$^4$  that these authors have determined for a Pop~III IMF with Salpeter slope with lower and upper end stellar masses of $M_{\rm *, lower}$ = 10 and $M_{\rm *, upper}$ = 100 M$_{\odot}$, which is broadly consistent with our 20 - 500 M$_{\odot}$ range.  These authors also report that for a choice of 100 - 500 M$_{\odot}$ the number of LW photons (per unit stellar mass) is lower by only a factor of two.  Therefore, while our choice is likely an overestimate, it can only be a slight overestimate.}
For Pop II stars we have used $\eta_{\rm LW}$ = 4000, which is consistent with Greif \& Bromm, as well as with Leitherer et al. (1999). 
In practice, the SFRs for each calculation are computed within the simulation at each timestep, and these are the values that
we use in equations (4) and (5) to determine the level of the LW background.

We note that we have taken a simplified approach to calculating the propagation of LW photons in which we assume that all LW photons generated by stars escape their host haloes, although it is likely that some fraction of LW photons are absorbed before escaping into the intergalactic medium (IGM) (see e.g. Kitayama et al. 2004; also Ricotti et al. 2001).
We have also not accounted for any metallicity dependence of the LW photon yield, as Leitherer et al. (1999) show this to be small (see their figures 75 and 77).

\subsubsection{The contribution from local sources}
While the LW background is a persistent source of H$_{\rm 2}$-dissociating photons, there are strong spatial and temporal 
variations in the LW flux that are produced locally by individual stellar sources (see e.g. Dijkstra et al. 2008; Ahn et al. 2009).
We account for these variations by tracking the formation and evolution of individual stellar clusters in our cosmological volume,
and by estimating the LW flux assuming a simple geometrical dilution of the photon density, whereby the LW flux goes as
$\propto$ 1/$r^2$, where $r$ is the distance from the stellar cluster.  As we are thus assuming both the IGM and the interstellar
medium (ISM) surrounding the stellar sources of LW photons to be optically thin\footnote{We do not, however, assume this when calculating the photodissociation rate of H$_{\rm 2}$, the self-shielding of which we treat 
as described in Section 2.2.3.}, our results are upper limits for the strength of the LW feedback.

In our implementation, the LW flux due to the $i$th individual stellar cluster of mass $m_{\rm *, i}$ (initially equal to an SPH particle mass of 10$^3$ M$_{\odot}$ in our star formation prescription), for the case of Pop~III and Pop~II star clusters, respectively, is

\begin{equation}
J_{\rm 21, III,*,i} \simeq 15 \left(\frac{r_{\rm i}}{1 \, {\rm kpc}} \right)^{-2} \left(\frac{m_{\rm *, i}}{10^3 \, {\rm M_{\odot}}} \right)  \mbox{\ ,}
\end{equation} 
and
\begin{equation}
J_{\rm 21, II,*,i} \simeq 3 \left(\frac{r_{\rm i}}{1 \, {\rm kpc}} \right)^{-2} \left(\frac{m_{\rm *, i}}{10^3 \, {\rm M_{\odot}}} \right)  \mbox{\ ,}
\end{equation}
where $r_{\rm i}$ is the distance to the $i$th cluster in physical coordinates.  We have arrived at this formula by taking the same number $\eta_{\rm LW}$ of LW photons per stellar baryon as described in Section 2.2.1, 
and by again assuming that these photons are produced at a constant rate over the 5 Myr maximum lifetime of the stars producing the bulk of the LW photons.

For every gas particle, we loop over the star particles contributing to the LW flux,\footnote{While we do this summation to find the total LW flux in the photon energy range 11.2 - 13.6 eV, we also account for the different spectra from Pop~II and Pop~III stars at energies $\ge$ 0.75 eV in calculating the total H$^-$ photodetachment rate, 
as explained in Section 2.2.4.} and sum up all of their individual 
contributions to find the total locally-produced LW flux, as follows:

\begin{equation}
J_{\rm 21, III, local} = \sum_{i=1}^{N_{\rm *, III}} J_{\rm 21, III,*, i} \mbox{\ ,}
\end{equation}
and
\begin{equation}
J_{\rm 21, II, local} = \sum_{i=1}^{N_{\rm *, II}} J_{\rm 21, II,*, i} \mbox{\ .}
\end{equation}
Here, for Pop~III and Pop~II stars, respectively, $N_{\rm *, III}$ and $N_{\rm *, II}$ are the total number of star particles within the simulation volume
with an age $\le$ 5 Myr (see e.g. Leitherer et al. 1999; Schaerer 2002).

\subsubsection{Self-shielding of H$_{\rm 2}$}
While we calculate the LW flux to which gas particles are exposed in the optically thin limit, assuming no attenuation due to absorption 
in the IGM or in the ISM surrounding the LW radiation sources, we do take into account the degree to which this flux 
is attenuated locally by the target gas due to self-shielding (e.g. Draine \& Bertoldi 1996; Glover \& Brand 2001).
To estimate the self-shielded LW flux we follow the approach suggested by Wolcott-Green et al. (2011).  In order to avoid prohibitively expensive computational routines, we choose an implementation which draws only on local quantities that are defined
for each SPH particle individually.  
We calculate the column density $N_{\rm H2}$ of H$_{\rm 2}$ over the local Jeans length, as follows:

\begin{equation}
   N_{\rm H2} = 2 \times 10^{15} \, {\rm cm}^{-2}  \, \left( \frac{f_{\rm H2}}{10^{-6}} \right) \left(\frac{n_{\rm H}}{10 \, {\rm cm}^{-3}} \right)^{\frac{1}{2}}  \left(\frac{T}{10^3 \, {\rm K}} \right)^{\frac{1}{2}}   \mbox{\ ,}
\end{equation}
where $f_{\rm H2}$ is the fraction of H$_{\rm 2}$, $n_{\rm H}$ is the number density of hydrogen nuclei, and $T$ is the gas temperature.  
Then, following the discussion in Wolcott-Green et al. (2011) on the formulae given in Draine \& Bertoldi (1996), the fraction $f_{\rm shield}$ by which the local LW flux is diminished due to self-shielding is estimated as 

\begin{eqnarray}
f_{\rm shield}(N_{\rm H2}, T) & = & \frac{0.965}{(1+x/b_{\rm 5})^{1.1}} + \frac{0.035}{(1+x)^{0.5}}  \nonumber \\
                   & \times & {\rm exp}\left[-8.5 \times 10^{-4} (1+x)^{0.5} \right] \mbox{\ ,} 
\end{eqnarray}
where $x$ $\equiv$ $N_{\rm H2}$/5$\times$10$^{14}$ cm$^{-2}$ and $b_{\rm 5}$ $\equiv$ $b$/10$^{5}$ cm s$^{-1}$.  Here $b$ is the Doppler broadening parameter, 
which for the case of H$_{\rm 2}$ absorbers is given by $b$ $\equiv$ ($k_{\rm B}$$T$/$m_{\rm H}$)$^{\frac{1}{2}}$, where $T$ is the temperature of the gas, $m_{\rm H}$ is the  mass
of atomic hydrogen, 
and $k_{\rm B}$ is the Boltzmann constant.  Thus, we have 

\begin{equation}
b_{\rm 5} = 2.9 \left(\frac{T}{10^3 \, {\rm K}}\right)^{\frac{1}{2}} \mbox{\ .}
\end{equation}
Then, we also have, using equation (10), 

\begin{equation}
x =  4 \left(\frac{f_{\rm H2}}{10^{-6}}\right) \left(\frac{n_{\rm H}}{10 \, {\rm cm^{-3}}}\right)^{\frac{1}{2}}  \left(\frac{T}{10^3 \, {\rm K}}\right)^{\frac{1}{2}} \mbox{\ .}
\end{equation}

With equations (12) and (13), the shielding factor $f_{\rm shield}$ (equation 11) is completely defined in terms of local quantities that can
be easily read from each SPH particle, making for a computationally inexpensive approach that nonetheless offers an estimate of 
the effect of self-shielding that is accurate to within $\simeq$ 15 percent (Wolcott-Green et al. 2011).

\subsubsection{The total dissociation and detachment rates}
To obtain the reaction rates that we use in our chemical network to account for the 
photodissociation of H$_{\rm 2}$ and the photodetachment of H$^-$, we add the 
cosmological background LW flux to the LW flux generated by local sources within
our simulation volume.\footnote{In principle, this leads to some double counting of
sources, but as we show in Appendix B the effect is only a modest overestimate
of the total LW flux.}  Given the fluxes (again in terms of $J_{\rm 21}$) calculated using equations (3), (4),
(7) and (8), we follow Shang et al. (2010) to obtain the corresponding photodissociation and
photodetachment rates.\footnote{We note that in using their formulae we have implicitly followed their assumption, roughly consistent with our modeling of these populations, that Pop II stellar surface temperatures are 10$^4$ K, while those of Pop III are 10$^5$ K.}  We thus find the following formula for the H$_{\rm 2}$ dissociation rate due to local sources:

\begin{eqnarray}
\lefteqn{k_{\rm H2,diss,local} = }  \\ 
&  &  13.6 \times 10^{-12} \, {\rm s}^{-1} \left(\frac{f_{\rm shield}}{0.1}\right) \left(\frac{m_{\rm *}}{10^3 \, {\rm M_{\odot}}} \right)  \nonumber \\
&  &  \times  \left[ \sum_{i=1}^{N_{\rm *, III}} \left(\frac{r_{\rm i}}{1 \, {\rm kpc}} \right)^{-2}   +   0.67  \sum_{i=1}^{N_{\rm *, II}} \left(\frac{r_{\rm i}}{1 \, {\rm kpc}} \right)^{-2}  \right] \nonumber \mbox{\ ,} 
\end{eqnarray}
where $f_{\rm shield}$ is again given by equation (10).
Also following the formulae from Shang et al. (2010), we have for the photodetachment rate of H$^{-}$, for which we neglect shielding due to its low abundance, 

\begin{eqnarray}
\lefteqn{k_{\rm HM,ion,local} = } \\
&  &  1.5 \times 10^{-10} \, {\rm s}^{-1}   \left(\frac{m_{\rm *}}{10^3 \, {\rm M_{\odot}}} \right) \nonumber \\
&  & \times \left[ \sum_{i=1}^{N_{\rm *, III}} \left(\frac{r_{\rm i}}{1 \, {\rm kpc}} \right)^{-2}   +  4 \times 10^3 \sum_{i=1}^{N_{\rm *, II}}  \left(\frac{r_{\rm i}}{1 \, {\rm kpc}} \right)^{-2}  \right] \nonumber  \mbox{\ .}
\end{eqnarray}

Similarly, we find the rate of H$_{\rm 2}$ dissociation due to the LW background radiation field as

\begin{eqnarray}
\lefteqn{k_{\rm H2,diss, bg} = } \\
&  &  1.4 \times 10^{-10} \, {\rm s}^{-1} \left(\frac{f_{\rm shield}}{0.1}\right) \left(\frac{1+z}{16}\right)^{3}  \nonumber \\
& & \times \left[\left(\frac{SFR_{\rm III}}{ {\rm M_{\odot} \, yr^{-1} \, Mpc^{-3}}}\right) 
+  0.67  \left(\frac{SFR_{\rm II}}{ {\rm M_{\odot} \, yr^{-1} \, Mpc^{-3}}}\right) \right]   \nonumber \mbox{\ ,} 
\end{eqnarray}
and we have for the rate of H$^{-}$ photodetachment due to the LW background radiation 

\begin{eqnarray}
\lefteqn{k_{\rm HM,ion, bg} =} \\ 
& & 1.5 \times 10^{-8} \, {\rm s}^{-1} \left(\frac{1+z}{16}\right)^{3}  \nonumber \\
& &  \times \left[ \left(\frac{SFR_{\rm III}}{ {\rm M_{\odot} yr^{-1} Mpc^{-3}}}\right)  +  4 \times 10^3 \left(\frac{SFR_{\rm II}}{ {\rm M_{\odot} yr^{-1} Mpc^{-3}}}\right)\right] \nonumber \mbox{\ .}
\end{eqnarray}

Finally, for each gas particle we add the contributions from both the background LW 
flux and the locally generated LW flux, such that the final rates
to appear in the chemical network for a gas particle are

\begin{eqnarray}
k_{\rm H2,diss,total} = k_{\rm H2,diss,bg} + k_{\rm H2,diss,local} \mbox{\ }
\end{eqnarray}
and
\begin{eqnarray}
k_{\rm HM,ion,total} = k_{\rm HM,ion,bg} + k_{\rm HM,ion,local} \mbox{\ ,}
\end{eqnarray}
where the rates appearing on the right hand side are given by equations (14), (15), (16) and (17).

\subsection{SN mechanical feedback and metal enrichment in the FiBY}
At the end of their brief lives, a large fraction of the the first Pop~III stars are expected
to explode as powerful SNe which eject the first heavy elements into the primordial gas.
This initiates the metal enrichment of the Universe which dramatically alters the 
cooling properties of the gas and which continues with stellar winds and SNe from all subsequent stellar populations.
We model this feedback following standard techniques that
have been developed by previous authors.  Here we briefly describe the implementations
that we have adopted for the chemical enrichment of the gas from stars and for 
the energetic feedback from SNe.  We refer the reader to the references provided for more details.

\subsubsection{SN mechanical feedback}
The mechanical feedback from SNe is modelled as a prompt injection of thermal energy into the ISM surrounding star particles 
(which represent individual, evolving stellar clusters), as described in Dalla Vecchia \& Schaye (2012).  For the feedback 
from Pop~II SNe, for each SN that occurs 10$^{51}$ erg of thermal energy is distributed stochastically to neighboring 
SPH particles by instantaneously assigning them a gas temperature 
of 10$^{7.5}$ K.  As Dalla Vecchia \& Schaye (2012) show, for the resolution of our simulations this prescription 
suffices to capture the deposition of mechanical energy into the ISM reliably well.  
We use the same technique to model feedback from Pop~III stars but we differentiate between type II  
SNe which occur for initial stellar masses 8 $\la$ $M_{\rm *}$ $\la$ 100, and the more powerful
PISNe which occur for initial stellar masses 140 $\la$ $M_{\rm *}$ $\la$ 260 (Heger et al. 2003).
For the former we inject 10$^{51}$ erg per SN, while for the latter we inject 3 $\times$ 10$^{52}$ erg 
per SN which is roughly the average PISN energy found from the suite of models computed by Heger \& Woosley (2002). 

\subsubsection{Metal enrichment}
We follow the prescription for metal enrichment presented in Wiersma et al. (2009a), which is
similar to that also employed by Tornatore et al. (2007b).  
In this implementation, Pop~II star particles continuously 
release hydrogen, helium, and metals into the surrounding gas in abundances calculated 
according to tabulated yields for types Ia and II SNe, and from asymptotic giant branch (AGB) stars.
The mixing of this material with the surrounding ISM is modelled by transferring it to neighboring 
SPH particles in proportions weighted by the SPH kernel.  
We use the same technique to model metal enrichment from Pop~III stars, but we adopt the appropriate different 
chemical yields for type II SNe (Heger \& Woosley 2010) and PISNe (Heger \& Woosley 2002).

\subsection{Reionization in the FiBY}
Concurrent with the build-up of the LW background radiation field is the onset of
reionization, the process by which the intergalactic medium becomes heated and
ionized at $z$ $\ga$ 6 (e.g. Ciardi \& Ferrara 2005).  
We adopt a simple approach to account for the effects of reionization.
In particular, we assume that reionization takes place uniformly throughout our simulation volume starting at $z$ = 12, 
roughly consistent with range of redshifts 
inferred for instantaneous reionization by WMAP (e.g. Komatsu et al. 2011) and also with the limit of $\Delta z$ $>$ 0.06 
for the extent of reionization reported by Bowman \& Rogers (2010).

In practice, to model the effects of reionization, at $z$ = 12 we switch from the collisional to photoionization equilibrium 
cooling tables, which account for heating by the ionizing background radiation field given by Haardt \& Madau (2001).  This results in a gradual heating of the IGM to $\sim$ 10$^4$ K.  To account for the shielding of dense gas from the
ionizing radiation, we adopt a maximum density threshold $n_{\rm shield,ion}$ below which the gas is subjected to the full radiative flux; following 
Nagamine et al. (2010) we choose $n_{\rm  shield,ion}$ = 0.01 cm$^{-3}$.  At densities above $n_{\rm shield,ion}$ the flux is decreased from the 
unattentuated value by a fraction ($n$/$n_{\rm shield,ion}$)$^{-2}$, which is proportional to the recombination rate and allows for a continuous
transition betwee the shielded and unshielded regimes.  The cooling rates are then derived by 
interpolation between the collisional equilibrium and photoionization equilibrium tables.

\begin{figure}
\includegraphics[width=3.4in]{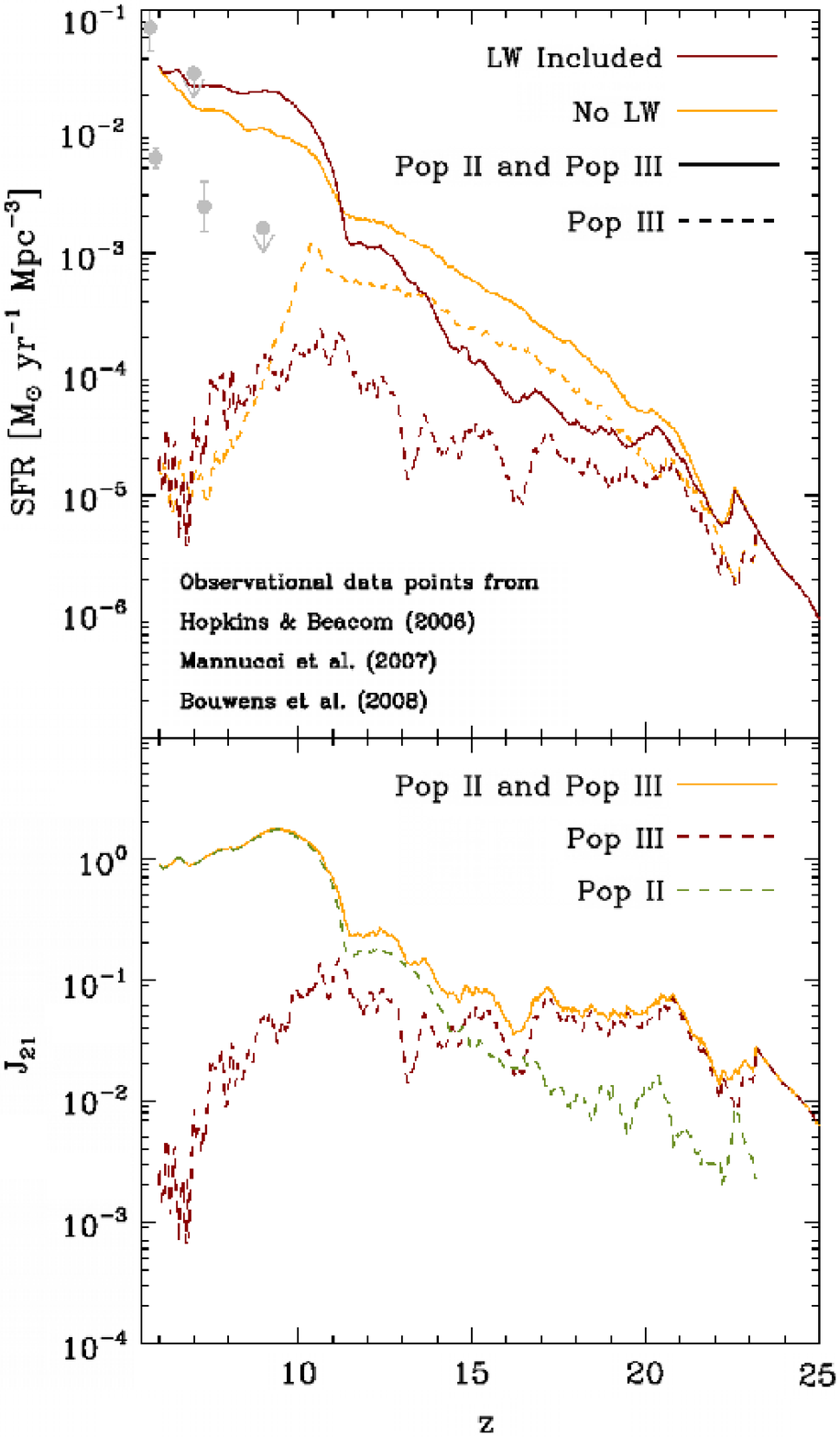}
\caption{{\it Top panel}: The comoving formation rate density of Pop III stars ({\it dashed lines}) and of 
all stars ({\it solid lines}), in our simulations with ({\it red}) and without ({\it yellow}) 
LW feedback, as a function of redshift $z$.  Also plotted here ({\it gray points}) are the star formation 
rate densities inferred from data on high-$z$ galaxies,  compiled from Hopkins \& Beacom (2006), Mannucci et al. (2007), 
and Bouwens et al. (2008).  {\it Bottom panel}: The level of the cosmological background LW flux in our simulation with LW feedback, 
in units of 10$^{-21}$ erg s$^{-1}$ cm$^{-2}$ Hz$^{-1}$ sr$^{-1}$, due to Pop II stars ({\it dashed green line}), 
Pop III stars ({\it dashed red line}), and both populations together ({\it solid yellow line}).} 
\end{figure}

\begin{figure*}
\includegraphics[width=6.9in]{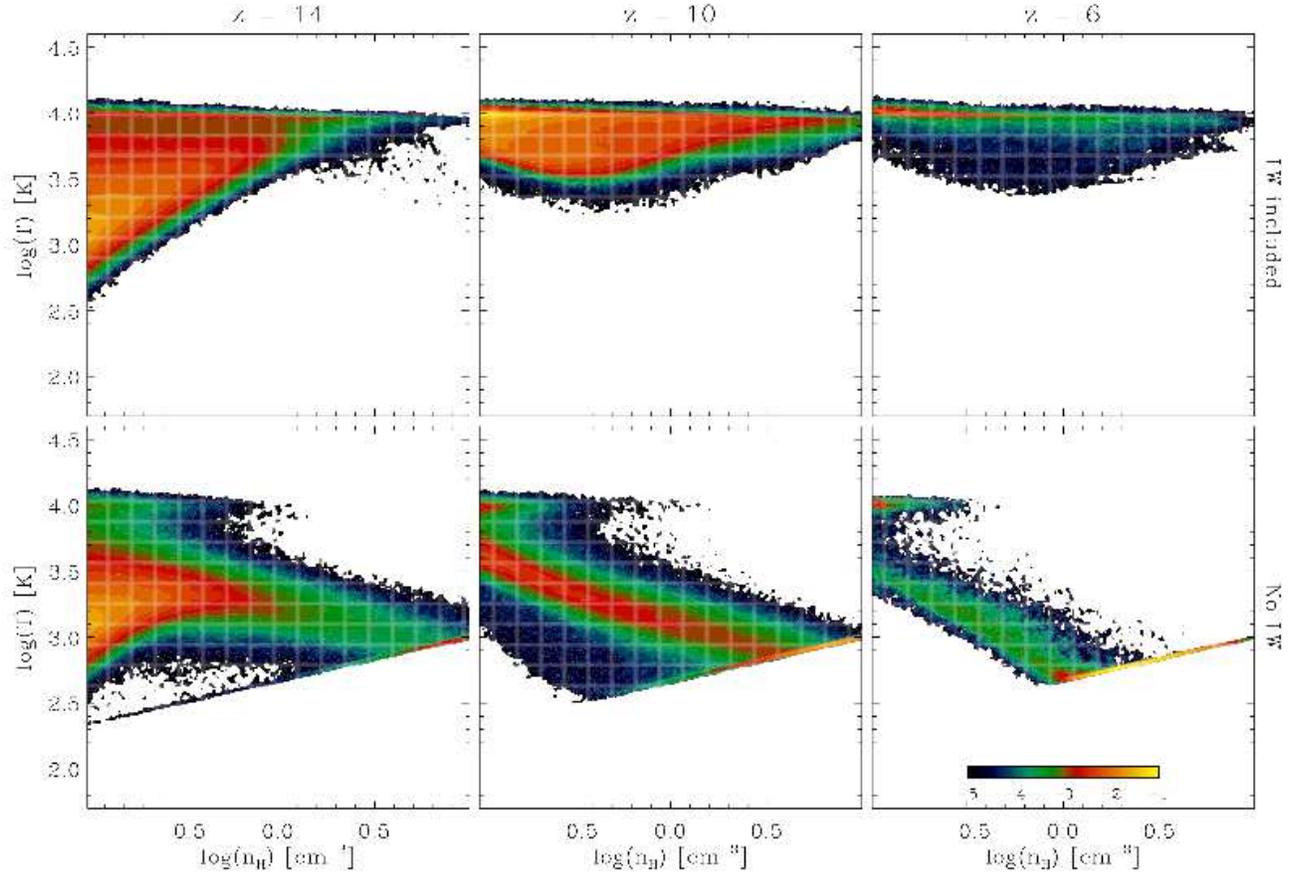}
\caption{The probability distribution function of primordial gas as a function of temperature $T$ and density $n$, in our simulations with LW feedback ({\it top}) and without it ({\it bottom}) at three representative redshifts: $z$ = 14 ({\it left panels}) , 10 ({\it middle panels}) and 6 ({\it right panels}).  The mass fraction of the gas is indicated by the contours, as shown by the scale in the bottom-right panel.  While the primordial gas collapses into haloes at typical temperatures $\sim$ 10$^3$ K at $z$ = 14, by $z$ = 6 the gas falls into to haloes having already been heated to $T$ $\sim$ 10$^4$ due to photoheating
and collisional heating.  Note that in the simulation including LW feedback the gas is hotter at the highest densities, due to the destruction of H$_{\rm 2}$ molecules that limits the efficiency with which the gas cools at $n$ $\ge$ 1 cm$^{-3}$. }
\end{figure*}

\section{Results}
Here we report the results of our two simulations, one including the effects of LW feedback (in dissociating H$_{\rm 2}$
and ionizing H$^-$, as described in Section 2.2) and one neglecting them.
We discuss a variety of results pertaining to the coevolution of the stellar Populations II and III that we model, as
well as implications for the detection of Pop~III PISNe and for the formation of black holes in the first galaxies.

\subsection{The global star formation rate}
The photodissociation of H$_{\rm 2}$ and the photodetachment of its intermediary
H$^-$ result in significantly decreased cooling efficiency of the primordial gas, 
which in turn reduces the Pop~III SFR and slows the process of chemical enrichment.  

The impact of this radiative feedback on the global (comoving) SFR density 
is shown in the top panel of Fig. 1, in which both the Pop~III and the total (Pop~II + Pop~III) SFRs are shown 
for both of our simulations.  The overall negative impact of the LW radiation is evident from 
the fact that both the Pop~III and the total SFRs are each lower in the simulation including
LW feedback, at redshifts $z$ $\ga$ 11.  Interestingly, however, this is not the case
at lower redshifts, but for different reasons for Pop~II and Pop~III.    Below we discuss
the evolution of the Pop~III and total SFRs separately.  As we shall argue, our results suggest
that the Pop~III SFR is regulated by LW feedback, while it is the pace of metal enrichment that largely
limits the Pop~II SFR.

\subsubsection{Lyman-Werner feedback and the Pop III SFR}
While the Pop~III SFR is similar in both simulations at the earliest times,
deviations begin to appear at $z$ $\simeq$ 20
when the LW background flux has reached a value of $J_{\rm 21,bg}$ $\simeq$ 0.05, as shown in Fig. 1.  This is consistent with the results of 
previous studies that have found that the cooling of the primordial gas becomes substantially reduced when
exposed to LW background fluxes of this order (e.g. Yoshida et al. 2003; Mesinger et al. 2006; 
Wise \& Abel 2007; O'Shea \& Norman 2008).

\begin{figure*}
\includegraphics[width=6.9in]{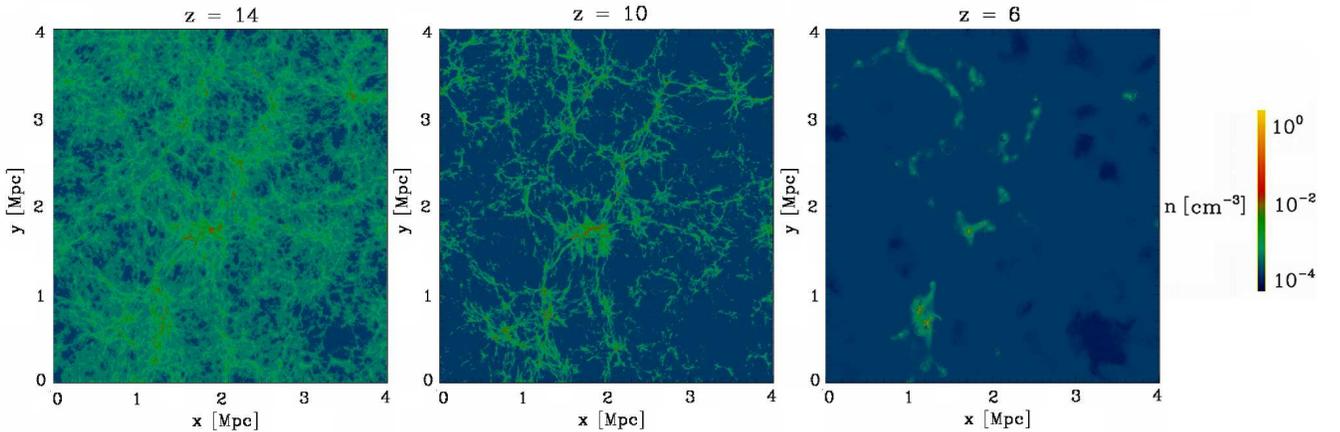}
\caption{The number density $n$ of the gas within a 400 kpc (comoving) slice of the simulation box, at redshifts $z$ = 14 ({\it left}), 10 ({\it middle}), and 6 ({\it right}).  The highest density regions host star formation which produces the metal-enriched outflows shown in Fig. 4, as well as  the LW radiation evident in Figs. 8 and 9.  }
\end{figure*}

\begin{figure*}
\includegraphics[width=6.9in]{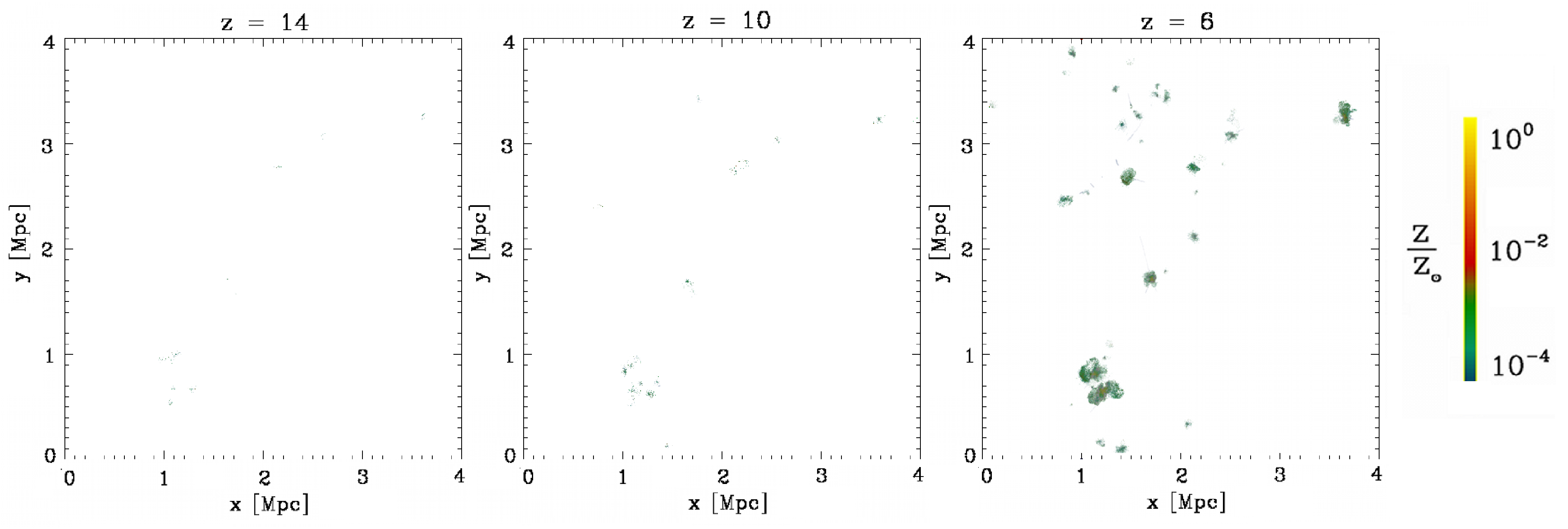}
\caption{The metallicity $Z$ of the gas relative to the solar metallicity $Z_{\odot}$ within the same 400 kpc (comoving) slice of the simulation box as shown in Fig. 3, at redshifts $z$ = 14 ({\it left}), 10 ({\it middle}), and 6 ({\it right}).  The white regions contain gas with metallicity $Z$ = 0.  The isolated nature of the metal-enriched regions is reflected in the small volume-filling fraction of metal-enriched gas shown in Fig. 5.}
\end{figure*}

The impact that the LW feedback has on the thermal properties of the gas is evident in Fig. 2, 
which shows the density and temperature of the primordial gas in each of the two simulations.  
In the simulation including LW feedback the gas is considerably hotter, due to the photodissociation 
of the H$_{\rm 2}$ molecules which provide cooling at densities $n$ $\ga$ 1 cm$^{-3}$.  
This less effective H$_{\rm 2}$ cooling translates into the lower Pop~III SFR in the simulation
with LW feedback, shown in Fig. 1.  The differences between the properties of the gas in the two
simulations grow with time, as both the background and local LW fluxes grow.  By $z$ = 6, the
primordial gas collapses into haloes at $\sim$ 10$^4$ K, due to the photoheating during reionization; 
while the gas cools to $T$ $\la$ 10$^3$ K with no LW feedback, it remains at $T$ $\sim$ 10$^4$ K
when it is included.  In principle, these higher temperatures may lead to the formation of
more massive Pop~III stars (e.g. O'Shea \& Norman 2008) or to the formation of supermassive stars
that collapse to for $\ga$ 10$^4$ M$_{\odot}$ black holes, as we discuss further in Section 3.4.

As Fig. 1 shows, the impact of LW feedback on the Pop~III SFR is fairly dramatic,
with the SFR varying by up to an order of magnitude between the two simulations, 
down to $z$ $\simeq$ 11.  This effect is of the same order as that predicted from simple modelling
of the build-up of the LW radiation field in the early Universe carried out by Johnson et al. (2008).  
Indeed, the reason for this similarity is likely due to the fact that the level of the LW background flux
we find is very close to the `critical' value of $J_{\rm 21,bg}$ $\simeq$ 0.04 that Johnson et al. (2008) 
argued should be generated in the early, Pop~III-dominated epoch.  The fact that we find slightly 
stronger suppression of Pop~III star formation than these authors 
predicted is likely due to our inclusion of Pop~II star formation, which produces a LW flux 
above that from just Pop~III stars which these authors considered (as shown in the bottom panel of Fig. 1).  
The agreement that we find with the results of Johnson et al. (2008) supports their claim that
LW background radiation regulates the Pop~III SFR such that the critical LW flux of $J_{\rm 21,bg}$ $\simeq$ 0.04
is produced.

This picture, however, breaks down with the onset of reionization feedback at $z$ $\la$ 12,
which has the effect of strongly suppressing Pop~III star formation by limiting the rate at which gas 
can cool and collapse into pristine DM haloes (e.g. Dijkstra et al. 2004; Johnson 2010).  While the Pop~III SFR drops by roughly 
an order of magnitude between $z$ = 12 and $z$ = 6, it is important to note that Pop~III star formation 
does persist down to this lower redshift even when accounting for the negative effects of 
LW feedback, reionization, and metal enrichment.  In fact, the Pop~III SFR that we find at $z$ $\sim$ 6
is very close to that found by Tornatore et al. (2007) in their simulations which accounted for reionization and
metal enrichment, but not LW feedback.  At higher redshifts, however, we find higher a Pop~III SFR than did 
these authors, likely due to the fact that the resolution of their simulation was not high enough to
resolve Pop~III star formation in the minihaloes in which it predominantly occurs at high redshift. 
We note that the SFRs we find are also in close agreement with those reported by Wise et al. (2012) 
from their simulations of galaxy formation in a smaller (1 Mpc)$^3$ comoving volume, which also accounted
for local LW feedback and metal enrichment, as well as the onset of reionization by local stellar sources.

\subsubsection{Metal enrichment and the Pop~II SFR}
As the primordial gas collapses into minihaloes and galaxies at high redshift, star formation and the 
concommitant metal enrichment of the gas soon follow.  This is evident in Figures 3 and 4, which show 
the density and the metallicity, respectively, of the gas in a 400 kpc (comoving) slice through our simluation volume
at three representative redshifts.  The densest regions in our simulated volume are also those which host Pop~III 
star formation and become
metal-enriched first, and as such they are the sites of the earliest Pop~II star formation.  As we discuss here,
the Pop~II SFR is governed largely by the rate at which metal enrichment occurs; in turn, this is dictated
by the LW feedback-regulated rate of Pop~III star formation.

The total SFRs found in each of our simulations are shown by the solid curves in the top panel of Fig. 1.  
The total SFR is lower at early times in the simulation with LW feedback, and the reasons
for this are at least two-fold.  Firstly, the lower Pop~III SFR due to the reduced cooling of the primordial 
gas partially accounts for the difference.  Secondly, though, the slower increase in the Pop~III SFR results 
in slower metal enrichment, and this in turn contributes to a slightly lower Pop~II SFR, since Pop~II stars
can only form in previously metal-enriched regions.  This second effect is also evident in Fig. 5, which shows the volume-filling fraction of metal-enriched gas in both simulations, as a function of redshift. At $z$ $\la$ 20, the delay in Pop~III star formation caused by the reduced fraction of H$_{\rm 2}$ in the simulation with LW feedback leads to a
slower rate of metal enrichment; in turn, this contributes to a reduction in the metal volume-filling fraction of up to an order of magnitude 
below that in the simulation without LW feedback.  

\begin{figure}
\includegraphics[width=3.4in]{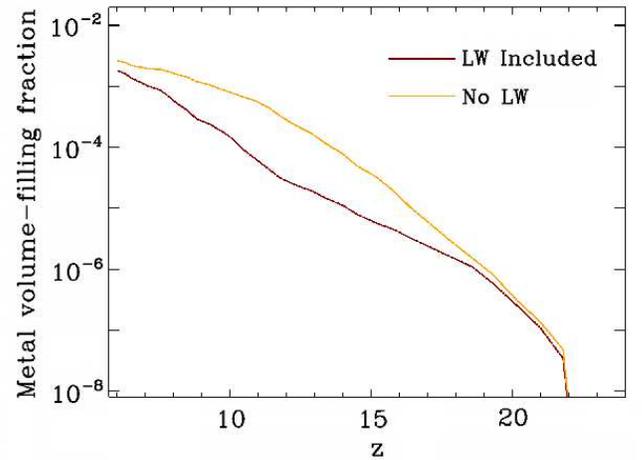}
\caption{The volume-filling fraction of gas enriched with metals, as a function of redshift $z$, in 
our simulations with ({\it red line}) and without ({\it yellow line}) LW feedback.}
\end{figure}

\begin{figure}
\includegraphics[width=3.4in]{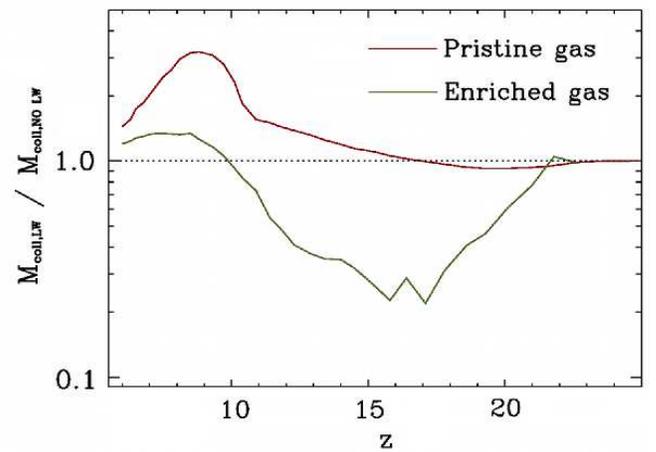}
\caption{The ratio of the mass $M_{\rm coll,LW}$  of pristine ({\it red line}) and metal-enriched ({\it green line}) of gas in collapsed haloes in the simulation with LW feedback to that 
($M_{\rm coll,NO LW}$) in collapsed haloes in the simulation without it.  The mass in collapsed pristine gas is almost always higher in the simulation with LW feedback, due to the
fact that the rate of Pop~III SNe is lower in this case and so less mass is blown out of haloes.  Because less gas is blown out in this case, despite the higher Pop~II SFR and SN rate
at  $z$ $\la$ 10, a larger mass of metal-enriched material remains collapsed in haloes at these redshifts.  It is because of this larger reservoir of collapsed metal-enriched gas that more
Pop~II stars form in the simulation with LW feedback, as shown in Fig. 7.}
\end{figure}

Interestingly, however, the metal volume-filling fractions in the two simulations begin to converge 
again at $z$ $\la$ 10.  This is partly due to the fact that the total SFR (and so also the metal production rate) in the simulation with 
LW feedback actually exceeds that in the simulation without it in this redshift range, as shown in Fig. 1.  
Due to the relatively low Pop~III SFR at these redshifts, it must be differences in the Pop~II SFR that drive 
this effect.  The likely cause of the enhanced Pop~II SFR is the increased minimum halo mass required for
star formation in a halo subjected to LW radiation (e.g. Machacek et al. 2001; O'Shea \& Norman 2008),
which results in star formation taking place in larger haloes in the simulation with LW feedback (see also Hummel et al. 2011; Wise et al. 2012).  As
dense star-forming gas is more readily retained in more massive haloes due to less efficient 
blow-away of the gas by SNe (e.g. Kitayama \& Yoshida 2005; Whalen et al. 2008), higher star formation
rates can be maintained in the more massive haloes that first form stars in the simuation with LW feedback.  
This is also consistent with the lower metal volume-filling fraction we find in the simulation with
LW feedback shown in Fig. 5, as it implies that less gas is blown out of haloes in this case.

Fig. 6 demonstrates that there is indeed a larger reservoir of metal-enriched gas in the simulation including LW feedback.  Shown is 
the ratio of the mass $M_{\rm coll,LW}$ of gas in collapsed haloes in the simulation with LW feedback to the ratio
of the mass $M_{\rm coll,NO LW}$ in the 
simulation without it.  The red line shows the ratio of pristine collapsed gas masses, while the green line 
shows the ratio of metal-enriched collapsed gas masses.  In the simulation with
LW feedback, the mass of collapsed pristine gas is almost always higher than that in the simulation without.  We attribute this
to the fact that, due to the lower Pop~III sar formation and SN rates with LW feedback on, less pristine gas is blown out of
collapsed haloes.  The mass of collapsed metal-enriched gas is at first much less in the case with LW feedback,
due to the decreased rate of Pop~III star formation and metal enrichment via SNe.  However, by $z$ $\la$ 10 this trend is 
reversed and the collapsed mass in metal-enriched gas becomes greater in the simulation with LW feedback.  Thus, 
there is indeed a larger reservoir of collapsed metal-enriched gas in this case, and this is consistent with the enhanced
Pop~II SFR at late times shown in Fig. 1.  Furthermore, the fact that the total collapsed mass is higher with LW feedback at the 
low redshifts at which the SFR is also higher strongly suggests that gas is more tightly bound in haloes due to LW feedback,
consistent with the explanation given above for the enhanced Pop~II SFR at late times.

\begin{figure}
\includegraphics[width=3.4in]{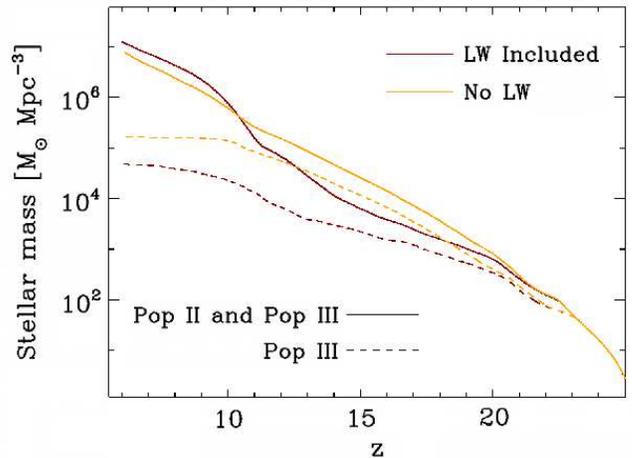}
\caption{The mass in Pop~III ({\it dotted lines}) and in both Pop~II and Pop~III ({\it solid lines}) stars formed by redshift $z$, in our simulations with ({\it red}) and without ({\it yellow}) H$_{\rm 2}$-dissociating LW feedback.  As expected, fewer Pop~III stars are formed in the simulation including LW feedback.  However, the total mass in stars formed is actually higher in the simulation with LW feedback.  As LW radiation delays the onset of star formation in low-mass DM haloes, they grow to larger masses before star formation begins.  In turn, as SN feedback is less efficient at blowing the gas out of more massive haloes, more second generation Pop~II star formation 
takes place in haloes when accounting for LW feedback.} 
\end{figure}

While the total SFR is increased by only a factor of $\la$ 2 at $z$ $\la$ 11 as a result of LW feedback, Fig. 7 shows that
this enhancement is strong enough that it has the effect of
increasing the total mass in stars that are formed at $z$ $\la$ 10. 
As this figure also shows, the mass in Pop~III stars is substantially lowered due to LW feedback, as expected from the reduction in 
the Pop~III SFR shown in Fig. 1.  Thus, the increase in stellar mass is entirely in metal-enriched Pop~II stars.
We note also that an added effect of the haloes retaining more gas in the case with LW feedback is 
that the gas is retained at higher densities and so is less susceptible to photoheating in our prescription for 
reionization.  This may also account in part for the higher Pop~II star formation rate at $z$ $\la$ 12, 
when reionization feedback is turned on.

\subsubsection{Comparison with observational data}
In the top panel of Fig. 1 we compare the SFRs that we find in our simluations to those
inferred from observations of high redshift galaxies, compiled from Hopkins \& Beacom (2006),
Mannucci et al. (2007), and Bouwens et al. (2008).  While our results are broadly consistent
with the SFRs reported by these authors, there are significant differences which may be due 
to the assumptions on the galaxy luminosity function and/or the stellar IMF which were 
made in order to estimate SFRs from the observations (for more discussion on this see Khochfar et al. 2012).  Despite these uncertainties in the 
modelling, that our results roughly agree with the observations provides a preliminary confirmation
of their validity.

\subsection{The epoch of reionization}
As discussed in Section 3.1.1, the dramatic drop in the Pop~III SFR at $z$ $\la$ 12 shown in Fig. 1 is due in 
large part to the onset of reionization.  Indeed, that this is not due solely to the LW feedback
is clear from the fact that the drop in the SFR is even more precipitous in the simulation neglecting LW feedback.  
As the impact of reionization in surpressing low-mass galaxy formation is evidently strong, we would like to 
verify that our modelling of this process has produced reasonable results.

As described in Section 2.4, we have modelled reionization in a simple way, by gradually ionizing and heating the 
gas in the IGM starting at redshift $z$ = 12.  While this approach is broadly consistent with 
the observational constraints on the epoch of reionization (e.g. Bowman \& Rogers 2010; Komatsu et al. 2009), 
we can check that this approach has also yielded results consistent with expectations.  As discussed by e.g. Pawlik et al. (2009),
one can estimate when the IGM is completely reionized by equating the rate at which ionizing photons from stars 
escape into the IGM with the rate at which the intergalactic gas recombines. 

If only a fraction $f_{\rm esc}$ of the hydrogen-ionizing photons produced by stars are able to escape due to the 
relatively high optical depth to photoionization, we  estimate the SFR required for reionization as

\begin{figure*}
\includegraphics[width=6.9in]{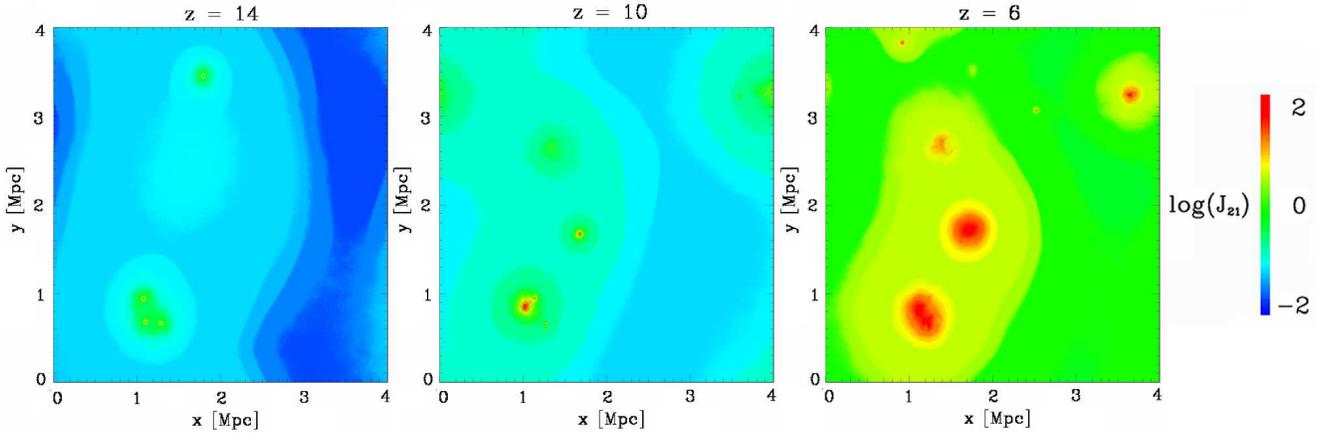}
\caption{The LW flux generated by all (Pop~II and Pop~III) stars within our simulation volume ($J_{\rm 21,local}$) to which gas is exposed in the same 400 kpc (comoving) 
slice of the simulation box as in the previous Figures, at the same three redshifts: $z$ = 14 ({\it left}), 10 ({\it middle}), and 6 ({\it right}).}
\end{figure*}

\begin{figure*}
\includegraphics[width=6.9in]{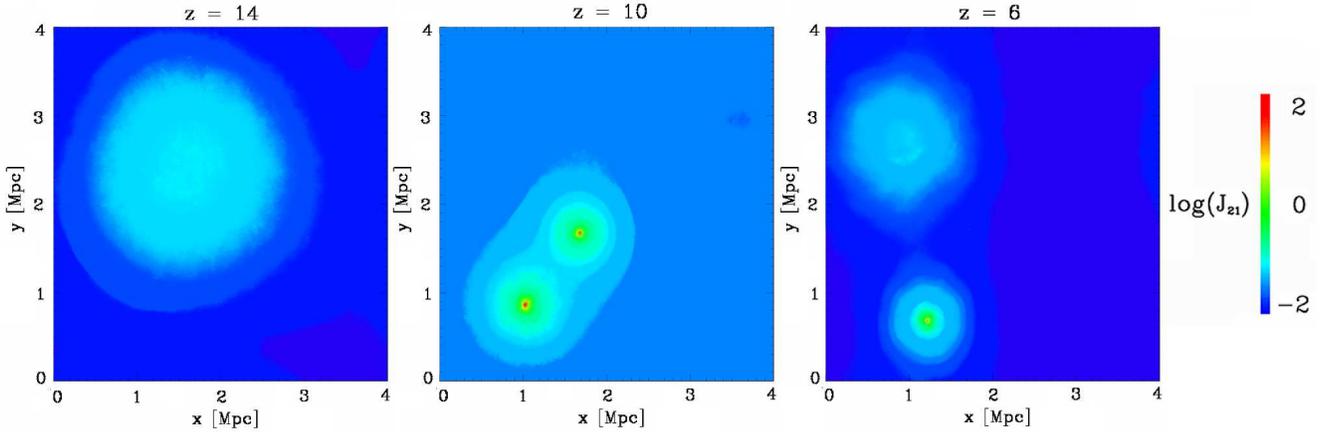}
\caption{The same as Fig. 8 but for the LW flux generated by only Pop~III stars ($J_{\rm 21,III,local}$).  While overall the LW flux produced by all stars steadily increases in time, as shown in Figs. 1, 8 and 10
the LW flux produced by Pop~III stars increases from $z$ = 14 to $z$ = 10, but then decreases from $z$ = 10 to $z$ = 6. }
\end{figure*}

\begin{eqnarray}
SFR_{\rm reion} & \simeq & 0.05 \, {\rm M_{\odot} \, yr^{-1} \, Mpc^{-3}} \,  \\ \nonumber
& \times & \left(\frac{C}{6} \right) \left(\frac{f_{\rm esc}}{0.1} \right)^{-1} \left(\frac{1+z}{7}\right)^3 \mbox{\ ,}
\end{eqnarray}

where we have normalized the clumping factor $C$ of the IGM to a typical value reported by Pawlik et al. (2009), 
and $f_{\rm esc}$ is normalized to a 
typical value found in cosmological radiative transfer simulations (e.g. Ricotti \& Shull 2000; Wise \& Cen 2009; Razoumov \& Sommer-Larsen 2010; Paardekooper et al. 2011; Yajima et al. 2011).  We note that this value of the required SFR may be an overestimate, as it was derived 
assuming that the stars producing ionizing photons are at solar metallicity.  Stars at lower metallicity, as would have formed
in the early Universe, produce more ionizing photons per stellar mass (e.g. Leitherer et al. 1999).

In our simulation we find that the total global SFR is $\simeq$ 0.03 M$_{\odot}$ yr$^{-1}$ Mpc$^{-3}$ at $z$ = 6, by which
time reionization is expected to be complete (see e.g. Ciardi \& Ferrara 2005).  
This is very close to the value of the global SFR required to achieve reionization, especially 
considering that $SFR_{\rm reion}$ in equation (20) is likely an overestimate.
This suggests that the SFR that we find in our simulation is sufficient to reionize the Universe
by $z$ $\sim$ 6 (see also Paardekooper et al. in prep).

\begin{figure*}
\includegraphics[width=6.9in]{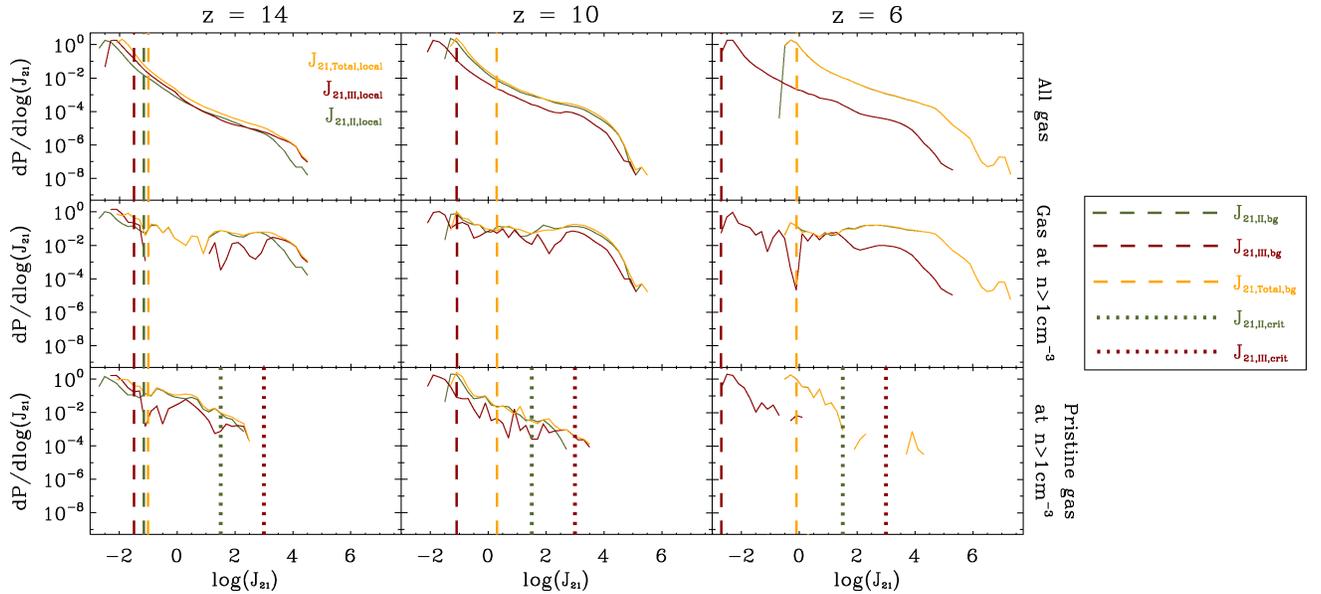}
\caption{The mass-weighted probability distribution functions (PDFs) of LW flux produced by individual (local) stellar sources.  The PDFs for all gas in the simulation ({\it top row}), just the densest ($n$ $\ge$ 1 cm$^{-3}$) gas ({\it middle row}), and just the primordial gas with $n$ $\ge$ 1 cm$^{-3}$ ({\it bottom row}) are shown, at $z$ = 6 ({\it right column}), 10 ({\it middle column}), and 14 ({\it left column}).
The PDFs of the flux generated by Pop~II stars only ({\it green}), Pop~III stars only ({\it red}), and the total flux generated by both stellar populations together ({\it yellow}) are shown in each panel.  The dashed vertical lines denote the level of the LW background radiation field produced by Pop~II stars only ({\it dashed red lines}), Pop~III stars only ({\it dashed green lines}), and both together ({\it dashed yellow lines}).  The dotted vertical lines in the bottom panels denote the critical values of the LW radiation field for the formation of black holes by direct collapse from dense gas in pristine haloes, if produced by Pop~II stars ({\it green dotted lines}) or by Pop~III stars ({\it red dotted lines}).  }
\end{figure*}

\subsection{Evolution of the sources of LW and reionizing radiation}
The bottom panel of Fig. 1 shows the contributions to the LW background both from Pop~III stars and
from all stars.  As the first Pop~III stellar populations begin forming and enriching the Universe as 
they evolve, Fig. 1 shows that they dominate the production of LW photons at redshifts down to $z$ $\simeq$ 15.  
At this point, however, the Pop~II SFR exceeds the Pop~III SFR enough to produce the majority of LW photons.  After
reionization sets in at $z$ $\la$ 12, the drop in the Pop~III SFR combined with the continued climb in the Pop~II SFR
results in the final shift to Pop~II stars producing nearly all of the LW radiation.  

As expected, the sources of the locally-generated LW flux ($J_{\rm 21,local}$) follow a similar pattern, as shown in Figures 8 and 9.  Overall, the total LW flux, from both Pop~II and Pop~III stars, increases with time from $z$ = 14 to $z$ = 6, as shown in Fig. 8; however, while
the LW flux from Pop~III stars increases from $z$ = 14 to $z$ = 10, it decreases from $z$ = 10 to $z$ = 6, as shown in Fig. 9.  This is due to 
the ever-increasing abundance of Pop~II star-forming haloes and the relative isolation and rarity of Pop~III star-forming haloes, which are
evident as the sources of the LW radiation in Figs. 8 and 9, respectively.

The transition between Pop~III and Pop~II sources of LW radiation is also evident in Fig. 10, which shows the mass-weighted
probability distribution functions (PDFs) of the LW flux $J_{\rm 21,local}$ from local stellar sources.\footnote{We note that in Fig. 10 we plot $J_{\rm 21}$ in the optically thin regime and have not accounted for the modest factor of $\la$  2 (see Wolcott-Green et al. 2011) 
decrease in $J_{\rm 21}$ at the highest densities we resolve ($n$ $\simeq$ 10 cm$^{-3}$) due to H$_{\rm 2}$ self-shielding (see Section 2.2.3); we do, however, account for self-shielding in computing the H$_{\rm 2}$ photodissociation rates that go into the chemical network that is solved in the simulation.}  The PDFs are shown at $z$ = 14, 10 and 6, and for gas in three different phases:  all gas in the simulation, all gas with density $n$ $\ge$ 1 cm$^{-3}$, and all metal-free gas with density $n$ $\ge$ 1 cm$^{-3}$.  At $z$ = 14, the PDF of the total LW flux largely follows the PDF of the LW flux from Pop~III 
stars, and the contribution from Pop~II stars is relatively small.  By $z$ = 10 the opposite is true, and by $z$ = 6 the PDF of the total LW flux is almost indistinguishable from the PDF of the flux from Pop~II stars.  This is consistent with our finding
that the Pop~II SFR begins to dwarf the Pop~III SFR at these redshifts.

We also note from Fig. 10 that the peak of the LW flux PDFs is generally at low values relative to the cosmological background LW flux levels ($J_{\rm 21, bg}$) that are denoted by the dashed vertical lines.  This  is consistent with our results suggesting that the Pop~III SFR density is largely regulated by the background radiation field (see Section 3.1).
That said, Fig. 10 also shows that for all redshifts and gas phases, the LW flux PDF exhibits a tail to high values of $J_{\rm 21, local}$.  Therefore, in rare regions the LW feedback is dominated by local stellar clusters and not by the LW background radiation field.  We discuss the implications of this further in Section 3.4.  

Because we calculate the LW flux due to local source as a function of the distance from the sources (see Section 2.2.2), the highest values of the locally-produced LW flux must be produced in the immediate neighborhood of star-forming regions, while the lowest values are reached in isolated regions far from star-forming regions.  Likewise, as we have a finite simulation volume, there is a minimum value of the flux from both stellar populations that can be reached in our simulation.  That the minimum flux from Pop~III stars is similar in each panel of Fig. 10, while that from the Pop~II stars increases to higher values over time, is a result of the relatively slow evolution of the Pop~III SFR and the much more rapid increase of 
the Pop~II SFR, as shown in Fig. 1.   Also, that the maximum value of the LW flux from Pop~II stars 
increases with time demonstrates that Pop~II star-forming regions become more clustered with time.  This is in contrast to the evolution of 
the maximum LW flux from Pop~III stars, which is nearly constant in time, demonstrating that Pop~III star-forming regions are not 
as clustered as Pop~II star-forming regions.  This is likely due to the effect of chemical enrichment, which occurs promptly after the
formation of Pop~III stellar clusters and shuts off subsequent Pop~III star formation by enriching the surrounding gas to metallicities above the 
critical value.

It is interesting to note that at $z$ $\ga$ 12, before reionization feedback is turned on in the simulation, Fig. 1 shows that
Pop~III stars account for a large fraction of the LW photons that are produced.  This fact, combined with 
the relatively high values for the escape fraction $f_{\rm esc}$ of ionizing photons expected for Pop~III star-forming haloes (e.g. Alvarez et al. 2006; Johnson et al. 2009) as compared to Pop~II galaxies (e.g. Gnedin et al. 2008), implies that the relative
contribution of Pop~III stars to reionization may have been relatively large.  However, Fig. 1 also shows that once reionization begins the negative feedback on Pop~III star formation can be very effective, and so we can not derive strong conclusions 
on the sources responsible for reionization without a simulation which fully couples the effects of reionization to the processes governing the Pop~III SFR.  Nevertheless, our results do suggest that Pop~III stars may have played a non-negligible role in beginning the process of reionization at $z$ $\ga$ 12.   

Previously, Dijkstra et al. (2008) and Agarwal et al. (2012) have also estimated the probability distribution of the H$_{\rm 2}$-dissociating flux 
in the early Universe (see also Ahn et al. 2009).   
While direct comparison with these two works is not straightforward since these authors 
present the flux PDFs of individual haloes while we show instead the flux PDFs of the gas both inside and outside haloes,\footnote{Direct comparison with Ahn et al. (2009) 
is also difficult, due to our different choices of simulation volume and resolution.}
we do note that our conclusion that local sources produce the highest fluxes was also found by these authors.  
We note also that the highest fluxes we find in pristine halos, shown in the bottom panels of Fig. 10, are broadly consistent with those found in these previous works.  
Next, we turn to discuss the implications of these plots, in particular.

\subsection{Implications for supermassive black hole seed formation}
One of the most promising scenarios for the formation of the seeds of supermassive black holes
in the early Universe is from the collapse of relatively hot $\sim$ 10$^4$ K primordial gas in a 
small fraction of the first protogalaxies (e.g. Bromm \& Loeb 2003; Begelman et al. 2006; Regan \& Haehnelt 2009).   At these high temperatures, the primordial gas is expected to collapse to a supermassive star that grows via
accretion to a mass $\ga$ 10$^4$ M$_{\odot}$  before collapsing to a black hole of similar mass (e.g. Shang et al. 2010; Johnson et al. 2012; see also e.g. Begelman 2010; Ball et al. 2011; Dotan \& Shaviv 2012; Hosokawa et al. 2012).  Once formed, these seed black holes may grow to become the supermassive black holes inferred to inhabit the centres of most galaxies today (e.g. 
Volonteri 2010).  In order for the primordial gas to be prevented from cooling to $\la$ 10$^4$ K via molecular transitions, it is likely 
that an elevated H$_{\rm 2}$-dissociating radiation field may be required to prevent the formation of 
molecules (e.g. Dijkstra et al. 2008; Shang et al. 2010; but see also Sethi et al. 2010; Inayoshi \& Omukai 2012).  

Shang et al. (2010; see also Omukai 2001; Wolcott-Green et al. 2011) find that the `critical' value $J_{\rm 21,crit}$ of the LW flux required in this scenario is 
sensitively dependent on the shape of the spectrum of the radiation, with the spectra of cooler stars producing more H$^-$-ionizing
radiation which prevents H$_{\rm 2}$ formation as discussed in Section 2.2.  For radiation produced by Pop~III stars with 
surface temperatures of 10$^5$ K and by Pop~II stars with surface temperatures of 10$^4$ K, they find critical
LW fluxes of $J_{\rm 21,III,crit}$ $\simeq$ 10$^3$ and $J_{\rm 21,II,crit}$ $\simeq$ 30, respectively.  
These values of the critical flux are shown in the bottom panels of Fig. 10, in which the PDFs of the LW flux for dense ($n_{\rm H}$ $\ge$ 1 cm$^{-3}$) primordial gas are presented.  As shown in the bottom-left panel, at $z$ = 14 a small fraction of the dense primordial gas (and an even smaller fraction of the overall gas) is exposed to a LW flux produced by Pop~II stars that exceeds the critical value for this Population.  
While not all of this gas is located in 
chemically pristine haloes that have not previously hosted star formation, we have verified that $\sim$ 10 percent of pristine gas 
does indeed reside in such haloes at $z$ = 14.  As it is likely that the gas must be of primordial composition in order for a supermassive star, and subsequently a black hole, to form (Omukai et al. 2008) we expect that it is in these haloes in which supermassive black hole seed formation
will occur.

This suggests that the conditions for direct collapse black hole formation may indeed be realized in rare regions where local stellar sources 
produce very high levels of LW flux.  These conditions are also found at redshifts down to at least $z$ = 6, as shown 
in the bottom-right panel of Fig. 10.  This is broadly 
consistent with the more detailed results found by Agarwal et al. (2012) on the frequency with which direct collapse
black holes form in the early Universe, and supports the view that a large fraction of supermassive black holes
may have originated via direct collapse (see also Petri et al. 2012).   

Finally, we note that this result suggests that this mechanism may have been the main avenue for black hole formation in the early Universe.
Even if all of the mass in stars formed by $z$ = 14 in our simulation, shown in Fig. 7, was converted into black holes, it would be less than the $\sim$ 10$^5$ M$_{\odot}$ expected for the mass of a black hole formed by direct collapse as outlined above (e.g. Shang et al. 2010; Johnson et al. 2012).  
Thus, it may be that most of the mass in black holes in the early 
Universe is contained in black holes formed by direct collapse in chemically pristine haloes exposed to high LW flux (see also Agarwal et al. 2012).

\begin{figure}
\includegraphics[width=3.4in]{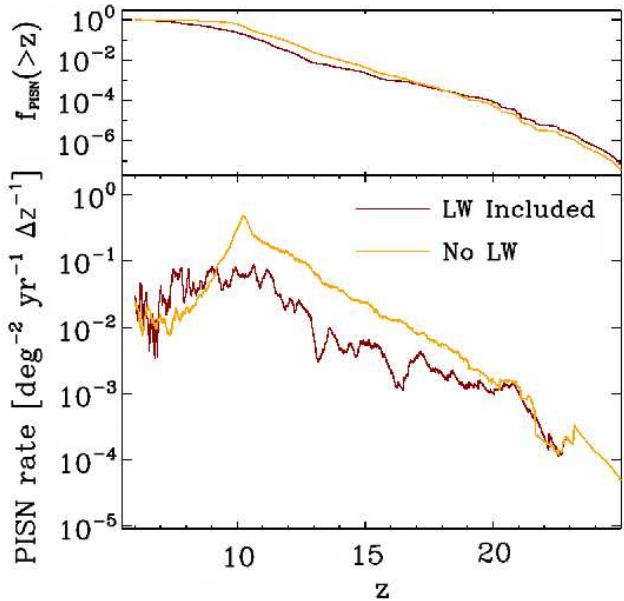}
\caption{The rate at which PISNe can be detected ({\it bottom panel}) from observations at $z$ = 0, as a function of $z$,
as inferred from our simulations with ({\it red curves}) and without ({\it yellow curves}) LW feedback.  Also shown is the cumulative fraction
$f_{\rm PISN}(> z)$ of PISNe detected at redshifts $>$ $z$ for observations over a given length of time ({\it top panel}).  
The effect of LW feedback is to reduce the PISN detection rate at most redshifts, simply due to the reduction in the Pop~III SFR shown in Fig. 1.  We find that the fraction of PISNe which could be detected at $z$ $\le$ 10
is substantially higher in the case with LW feedback.   }
\end{figure}

\subsection{The expected Pop III pair-instability supernova rate}
It is likely that Pop~III star-forming galaxies in the early Universe are too dim to be identified in upcoming surveys 
by even the most powerful telescopes, such as the {\it James Webb Space Telescope} (JWST) (Barkana \& Loeb 2000; Gardner et al. 2006; Ricotti et al. 2008; Johnson et al. 2009; Pawlik et al. 2011).\footnote{There is, however, the possibility of observing Pop~III galaxies if they are gravitationally lensed (see e.g. Zackrisson et al. 2012).}  However, if the Pop~III IMF is sufficiently top-heavy, there is a good possibility of detecting
Pop~III pair-instability supernovae (PISNe) from primordial galaxies (e.g. Scannapieco et al. 2005; Weinmann \& Lilly 2005; Wise \& Abel 2005; Mesinger et al. 2006b; Hummel et al. 2012; Pan et al. 2012), given their high luminosities and spectral signatures (e.g. Kasen et al. 2011; Frey et al. 2012).

As our simulations go beyond previous work in modelling the extended epoch of Pop~III star formation, it is 
of interest to estimate from them the rate at which primordial PISNe are produced.  The bottom panel of Fig. 11
shows the rates of PISNe as seen on the sky that we find from our simulations.  We have arrived at this rate following
the approach of Bromm \& Loeb (2006), and assuming our Pop~III stellar IMF (see Section 2.1.1) and a stellar mass range 
for PISNe of 140 - 260 M$_{\odot}$ (see e.g. Heger et al. 2003).\footnote{As found by Chatzopoulos \& Wheeler (2012), 
significantly less massive Pop~III stars may produce PISNe if they rotate rapidly. Changing the minimum mass for PISN to that which 
these authors find ($M_{\rm *}$ $\simeq$ 65 M$_{\odot}$) in our calculation would increase the PISNe rates we find by a factor of $\simeq$ 4.}  
Comparing our results to those presented in the recent work by Hummel et al. (2012), 
we find rates that are roughly an order of magnitude lower than the upper limit these authors estimate, likely due in 
large part to the less top-heavy Pop~III IMF that we have adopted.\footnote{Our results are also in rough agreement with
those of Wise \& Abel (2005) at high redshifts ($z$ $\ga$ 20), after acccounting for our different choices of cosmological parameters.} 
Therefore, we find agreement with their conclusion that searches for PISNe by the JWST will likely have to be conducted over many fields of view if even one is to 
be discovered.  

This result notwithstanding, such surveys will be instrumental for testing predictions such as those 
we present here based on detailed numerical simulations.  One prediction that suggests 
Pop~III PISNe will be found, if they are produced at a sufficiently high rate, is that PISNe are most likely to occur
at relatively low redshifts.  The top panel in Fig. 11 shows the fraction of PISNe $f_{\rm PISN}$($z$) 
that originate at redshifts $>$ $z$.  Interestingly, the effect of LW feedack is to increase the fraction of PISNe
that occur at $z$ $\la$ 10 from $\simeq$ 50 in the case without LW feedback to $\simeq$ 80 percent.
Therefore, the vast majority of Pop~III PISNe may occur at much lower redshifts than
those at which the first stars typically form ($z$ $\sim$ 20), and may therefore be more easily 
detectable given their closer proximity.  This follows directly from our finding that Pop~III star formation
continues down to at least $z$ = 6, despite the strong radiative, chemical, and mechanical feedback 
exerted by both primordial and metal-enriched stars in the early Universe.

\section{Discussion and Conclusions}
We have presented the results of large-scale cosmological simulations of the
earliest stages of galaxy formation, in which the feedback from both Pop~II and
III stars plays a defining role.  In modelling the mechanical, chemical, and 
radiative feedback processes that regulate the formation of both of these
stellar populations, we have tracked their coevolution in a self-consistent
manner.  

We have found that in the early Universe the Pop~III star formation rate is largely 
regulated by the global LW background, while Pop~II star formation rate is instead largely 
regulated by the pace of metal enrichment.  As these feedback processes are linked,
the two Populations coevolve in a complex manner.
In particular, we find that the Pop~III SFR is regulated to be at a level very close to 
that at which LW feedback becomes effective in suppressing the cooling of primordial gas
in minihaloes ($J_{\rm 21,bg}$ $\simeq$ 0.04), as also found previously in much more simplified 
simulations (e.g. Johnson et al. 2008). The total SFR is also decreased due to LW feedback at early times,
both due to the lower Pop~III SFR and to the slower pace of chemical enrichment,
which by definition must precede Pop~II star formation.  At later times ($z$ $\la$ 11),
however, most likely due to the decreased mechanical feedback from SNe in blowing away 
the gas in relatively small DM haloes, we find that the effect of LW radiation is to 
raise the total SFR by a factor of $\la$ 2 above the rate obtained in its absence.
This leads to the counter-intuitive result that the total mass in stars formed by
$z$ $\sim$ 6 is in fact increased due to LW radiation.

While all of the feedback effects that we have included are expected to 
negatively impact the rate of Pop~III star formation, we find that down to $z$ $\sim$ 6
Pop~III stars still form at a rate per comoving volume of $\sim$ 10$^{-5}$ M$_{\odot}$ yr$^{-1}$ Mpc$^{-3}$,
just one order of magnitude below its peak value at $z$ $\sim$ 10.  This continuation of Pop~III star formation
down to such low redshifts implies that $\sim$ 80 percent of primordial PISNe occur at $z$ $\la$ 10.  However, 
we have also confirmed that their overall low rate of occurence will likely require many 
fields of view to be surveyed by the JWST in order even a single PISN to be discovered.

While our simulations are some of the largest and most comprehensive
to date, we have not fully self-consistently tracked every important physical 
process governing galaxy formation.  In particular, we have taken a simplified 
approach to account for the impact of reionization on the heating of the IGM, and 
we have neglected the impact of ionizing radiation on the ISM surrounding 
stellar clusters.  Nonetheless, we find that our simple assumption of global reionization 
beginning at $z$ = 12 yields a global SFR that is sufficient to completely reionize 
the IGM by $z$ $\simeq$ 6.  We note, however, that not accounting for the photoionization of the ISM within 
star-forming haloes, especially in the first Pop~III star-forming haloes, likely 
makes our results lower limits for the efficiency of metal enrichment; as shown 
by e.g. Kitayama \& Yoshida (2005) and Whalen et al. (2008), photoheating
lowers the density of the ISM and allows SN ejecta to more readily escape into the IGM.
Incidentally, because this photoheating of haloes by internal stellar sources  
leads to more gas being blown out of low-mass Pop~III star-forming haloes,
accounting for it would also likely result in a further enhancement of 
Pop~II star formation at late times in the case with LW feedback.

Finally, because we have modelled the spatial and temporal variation of the LW 
radiation generated by stars, together with metal enrichment, in a sufficiently 
large cosmological volume, we are able to identify regions of dense, primordial
gas exposed to very high levels of LW radiation.  We find that such regions 
exist at $z$ $\ga$ 6 in our 64 Mpc$^3$ (comoving) cosmological volume.
This result corroborates other recent work (e.g. Agarwal et al. 2012) in supporting the theory 
that these sites may play host to the formation of the supermassive ($\ga$ 10$^4$ M$_{\odot}$)
stellar seeds of the black holes inhabiting the centres of galaxies today.

To summarize, our most important results are the following:

\begin{itemize}

\item Despite the negative feedback from LW radiation, chemical enrichment and photoionizing radiation
during reionization, significant Pop~III star formation continues down to at least $z$ $\simeq$ 6 (see Section 3.1.1).

\item  We find that LW feedback leads to an overall enhancement 
in Pop~II star formation, as compared to the case without LW feedback. 
We attribute this to the fact that Lyman-Werner feedback delays the
onset of Pop~III star formation until haloes are larger and less susceptible 
to gas blow-out by SNe, which results in larger reservoirs of gas for Pop~II star formation (see Section 3.1.2).

\item Sufficiently high LW fluxes are produced for the primordial gas to collapse 
into $\sim$ 10$^5$ M$_{\odot}$ black holes by direct collapse, even within 
our relatively small (4 Mpc)$^3$ comoving simulation volume (see Section 3.4).

\item Due to the continuation of Pop~III star formation down to $z$ $\simeq$ 6, a majority of the primordial 
PISNe that may be detected in upcoming surveys will likely be found at $z$ $\la$ 10 (see Section 3.5).

\end{itemize}

We have focused in the present work on the role of the LW radiaton produced by 
stars in governing the global SFR and metal enrichment in the early Universe. 
Many other issues, including the progress of chemical enrichment and 
the transition from Pop~III to Pop~II star formation, the global and individual 
properties of galaxies and DM haloes formed by $z$ $\sim$ 6, and the role of star-forming galaxies in
the inhomogeneous process of reionization, will be explored in greater detail in additional work
in the FiBY project.

\section*{Acknowledgements}
JLJ gratefully acknowledges the support of the U.S. Department of Energy through the LANL/LDRD Program.  CDV acknowledges support by Marie Curie Reintegration
Grant FP7-RG-256573.  We acknowledge helpful discussions with Bhaskar Agarwal, Volker Bromm, Umberto Maio, Jan-Pieter Paardekooper and Dan Whalen, and we thank Andrew Davis for comments on an early draft.


\appendix

\section{Resolving LW feedback}
In order to capture the impact of H$_{\rm 2}$ photodissociation on the thermal properties of the gas, we must resolve gas above a certain 
minimum density at which H$_{\rm 2}$ cooling is effective.

This minimum density can be estimated by noting that at high redshifts the primordial gas lies on the adiabat passing through $T$ $\simeq$ 2000 K and $n$ $\simeq$ 1 cm$^{-3}$, which are typical
values found for the gas collapsing into minihaloes when it first begins cooling via H$_{\rm 2}$ (e.g. Yoshida et al. 2003; Johnson et al. 2008).  
If H$_{\rm 2}$ cooling is not effective, the gas will remain on this adiabat until it collapses to densities at which atomic hydrogen cooling is effective.  

Noting that atomic hydrogen cooling is important at temperatures $\ga$ 10$^4$ K, we can solve for the density at which the
adiabatic gas will begin cooling by atomic hydrogen emission.  As in this case the temperature evolves according to $T$ $\propto$ $n^{\frac{2}{3}}$,
we find this density to be $n$ $\simeq$ 10 cm$^{-3}$.  Thus, at densities $n$ $\ga$ 10 cm$^{-3}$ gas can cool and collapse even in the absence of 
H$_{\rm 2}$, as atomic hydrogen cooling alone will maintain it in a roughly isothermal state at $T$ $\simeq$ 10$^4$ K (e.g. Haiman 2009).
However, at densities $n$ $\la$ 10 cm$^{-3}$ the gas can only cool effectively via H$_{\rm 2}$ molecules.  

Therefore, in our simulation we have chosen $n$ = 10 cm$^{-3}$ as the threshold density for star formation, such that the evolution of the gas is properly resolved 
up to this density.  This ensures that we resolve the impact of LW feedback in diminishing the efficiency of gas cooling at the densities at
which the collapse of the gas relies most heavily on H$_{\rm 2}$ cooling.

\section{Double counting sources}
As our estimates of both the cosmological background LW flux and the flux generated locally by individual sources
are based on the star formation rate within our simulation volume, it is important to check that any error
invoked by double counting these sources is small.  To address this, we note that the background radiation field is predominantly produced by sources outside of the simulated volume. 
To see this, we can express the background flux as the integral over all sources within the LW horizon $r_{\rm LW}$, the maximum distance which LW photons can travel, 
which is typically $r_{\rm LW}$ $\la$ 10 Mpc (physical) (e.g. Haiman et al. 1997).
For a cosmological stellar mass density $\rho_{\rm *}$ and a flux from individual stars $J_{\rm 21,*}$ $\propto$ $r^{-2}$, we have for the background flux 

\begin{equation}
J_{\rm 21, bg} \propto \int^{r_{\rm LW}}_{\rm 0} 4\pi r^2 \rho_{\rm *} J_{\rm 21,*} dr  \propto r_{\rm LW} \rho{\rm *} \mbox{\ .}
\end{equation}
Therefore, the level of the background flux is proportional to the LW horizon $r_{\rm LW}$.  A simple estimate of the fractional error introduced by double counting sources is given 
by the ratio of our box length (4 Mpc comoving) to $r_{\rm LW}$.  As a function of $z$, this is only $\sim$ 0.04 [(1+$z$)/10]$^{-1}$, assuming $r_{\rm LW}$ = 10 Mpc (physical).  

Therefore, the stars in our simulation volume produce only a small fraction of the background radiation field, and the 
error in using them to compute the background LW flux in addition to the local flux should be small.  This is verified by the fact that the local LW flux PDFs in Fig. 10 show
that most of the gas is exposed to local LW fluxes $J_{\rm 21,local}$ lower than the cosmological background flux.

\end{document}